\documentstyle[prl,aps,twocolumn,psfig]{revtex} 
\begin{document}
\draft
\date{
\today}
\title{Interaction-assisted propagation of Coulomb-correlated
electron-hole pairs in disordered semiconductors}
\author{D.~Brinkmann$^1$, J.E.~Golub$^1$, S.W.~Koch$^1$, 
P.~Thomas$^1$, K.~Maschke$^2$,
I.~Varga$^{1,3*}$}
\address{$^1$Fachbereich Physik und wissenschaftliches Zentrum f\"ur
Materialwissenschaften,
Philipps-Universit\"at Marburg,
Renthof 5, D-35032 Marburg, Germany, \\
$^2$Institut de Physique Appliqu\'ee,
\'Ecole Polytechnique F\'ed\'erale de Lausanne, CH-1015 Lausanne,
Switzerland \\
$^3$Institut f\"ur Theoretische Physik, Universit\"at zu K\"oln,
Z\"ulpicher Stra\ss e 77, D-50937 K\"oln, Germany}

\maketitle

\begin{abstract}
A two-band model of a disordered semiconductor is used
to analyze dynamical interaction induced weakening of localization
in a system that is accessible to experimental verification.
The results
show a dependence on the sign of the two-particle
interaction and on the optical excitation energy 
of the 
Coulomb-correlated electron-hole pair.
\noindent
\end{abstract}
\narrowtext

The problem of two interacting particles (TIP) in a random potential
is an excellent paradigm for the general question of the interplay of
disorder and interactions in many-body systems.  First addressed in
a 1990 paper by Dorokhov \cite{dho}, the subject has been especially
well studied since Shepelyansky's 1994 publication \cite{Shepel}.
Considering the TIP localization length $l_2$, most authors
\cite{Shepel,Imry,Jaquod,Jaquod1,frahm} obtain an interaction-induced
increase $l_2 > l_1$ over the single-particle localization length
$l_1$ independent of the sign of the interaction, with $l_2/l_1 \sim
l_1^a$ and $a=1$ or $a=0.65$ \cite{frahm}.  Here, $l_{1}$ and $l_{2}$
are measured in units of the lattice constant of a one-dimensional
Anderson chain.  Similar results have been obtained for TIP in a
quasiperiodic chain \cite{evan}.  The independence of the predicted
effect on the sign of the interaction is an especially intriguing
feature.

Early work approached this problem using a wide variety of theoretical 
techniques
and focused on establishing the existence of the TIP-effect, while
more recent work
has dealt with quantitative details like scaling behaviour and the
influence of interaction range, strength and sign
\cite{dho,Shepel,Imry,Jaquod,Jaquod1,frahm,frahm1,oppen,Roemer,Ponomarev}.

Existing works comprise purely theoretical case studies since
the model of just two particles in a single band does not
correspond to any real physical situation.  Obviously, experimental
study is needed to promote
further understanding of the TIP problem,
and would put the presently rather academic discussion on a firm
physical basis.  Exploiting the fact that the coherent spatio-temporal
dynamics of Coulomb-correlated electron-hole pairs is strongly
influenced by the two-particle interaction,
we show in the present
paper that {\it the TIP localization properties should be accessible to
modern ultrafast optical techniques}.  The corresponding spatial and
temporal scales are on the order of sub-$\mu$m and 1 ps. 
Our numerical studies are based on integration of the
Semiconductor Bloch Equations and include no {\it a priori}
assumptions about energy hierarchies or interaction matrix elements.
The calculations were performed for a disordered 1D semiconductor
quantum wire, where localization effects are most important.  Despite
its simplicity, this model system contains already all essential
ingredients to describe the dynamics following optical excitation even
in disordered systems of higher dimension \cite {foot}.  The model
parameters have been given values that resemble those of
realistic disordered semiconductor quantum wires.  Additional physical
parameters (excitation energy, spectral pulse width, screening length,
different masses of the two particles) allow the study of a wide
variety of observable phenomena.

In our numerical calculations we investigate the spreading of an
electron-hole wave packet after local excitation by an optical
pulse. Here the interaction is given by the long-range Coulomb
potential which, besides producing bound states (excitons) near the
edges of the excitation spectrum, also correlates the electrons and
holes in the pair continuum. Previous theoretical studies of the
spatial-temporal dynamics of wave packets formed from excitons show
that their motion is rather limited in the presence of scattering
\cite{stein}. This result is recovered by our present calculations.
Here we focus our interest on the dynamics of optically generated wave
packets in the pair continuum.  We find that the excitation conditions
in the presence of particle-particle interaction influence the carrier
dynamics dramatically. In addition, and in contrast to some previous claims
in the literature, we find that the sign of the interaction has a
pronounced effect on the spatio-temporal dynamics.


We consider a 1D array of sites $i$ with diagonal disorder in both the
valence band ($vb$) and the conduction band ($cb$). The site energies
$\varepsilon_{vi}$ and $\varepsilon_{ci}$ corresponding to the $vb$
and the $cb$, respectively, are randomly distributed over the interval
$\left[-W/2,W/2 \right]$ and are uncorrelated.  The nearest neighbor $cb$
levels are coupled by the tunneling term $-J^c$, the $vb$ levels by
$-J^v$.  We use the Coulomb interaction in its monopole-monopole form
\cite{stahl} with matrix elements
\begin{equation}\label{coul}
V_{ij} = -\frac{U}{4\pi\epsilon\epsilon_0}\frac{e^2}{r_{ij}+\alpha}
\end{equation}
which has been regularized in order to cope with the pathological
singularity in one dimension. The constant $\alpha$ has been chosen to
be 5 times the lattice constant \cite{banyai}.  The total Hamiltonian
is written as $\hat{H}=\hat{H_0}+\hat{H_I}+\hat{H_C}$, where
$\hat{H_0}$ describes the $vb$ and $cb$ band structure, $\hat{H_I}$
represents the 
semiclassical interaction with the electromagnetic field $E_i(t)$ in
dipole approximation, and $\hat{H_C}$ defines the electron-electron
interaction term \cite{klaus,dirk}.

In the following, we assume a \em local\em\/ initial excitation at the
central site $i=0$, which is modeled by setting $\mu_i = \delta_{i,0}$
for the local dipole matrix element in $\hat{H_I}$.  The optical
polarization $p_{ij}(t)$ is obtained from the equation of motion for
the polarization operator $\hat{p}_{ij}=\hat{d}_i \hat{c}_j$, which is
coupled to the equation of motion of the electron and hole intraband
quantities $\hat{n}^e_{ij}=\hat{c}_i^{+} \hat{c}_j$ and
$\hat{n}^h_{ij}=\hat{d}_i^{+} \hat{d}_j$, respectively, where the
operators $\hat{c}_i^+,\hat{c}_i$ ($\hat{d}_i^+,\hat{d}_i$) describe
the electron (hole) creation and annihilation operators at site $i$.
The equation of motion for the expectation values $p_{ij}(t)$ and
$n^{e,h}_{ij}(t)$ is treated 
using 
the well-known Semiconductor Bloch Equations
for $p_{ij}(t)$ and $n_{ij}^{e,h}(t)$ in the real-space representation 
\cite{dirk}.
Detailed derivations of the Semiconductor Bloch Equations are given 
in \cite{stahl} and the textbook \cite{haug}.  
As we are interested in small excitation
densities, we write only the equation for $p_{ij}$ in the lowest
(linear-response) order in the exciting field
\begin{eqnarray} 
\partial_t \,p_{ij} =&-& i
\Bigl(\varepsilon^e_i - \varepsilon^h_j - V_{ij}\Bigr) p_{ij}
  + i \sum_{l=1}^{N} 
  \left( J^e \,p_{il} 
 + J^h \,p_{lj}  \right)\nonumber\\
& + & i 
\mu_j E_j(t)\, \delta_{ij}.
\end{eqnarray}
Using the conservation
laws
%
$ n_{ij}^e = \sum_l p_{lj} \, p_{li}^*$ and
$ n_{ij}^h = \sum_l p_{jl} \, p_{il}^*$
%
valid in this lowest order \cite{wer}, we obtain the intraband
quantities.

Instead of studying a rather academic localization length which
describes only the asymptotic behavior of wave functions, we calculate
the experimentally more relevant participation number
%
$\Lambda (t) = (\sum_i n_{ii}^2)^{-1}$.
%
Here $n_{ii}$ stands for either $n_{ii}^e$ or $n_{ii}^h$.  With the
packet localized at site $0$, $n_{ii}= \delta_{i0}$ and $\Lambda=1$,
while for an excitation uniformly extended over the sample of $N$
sites, $n_{ii}=1/N$ and $\Lambda=N$.  Our calculations were performed
for chains containing $N=240$ sites.  Boundary effects can easily be
identified in the temporal evolution of $\Lambda$
and do not play any role as long as
$\Lambda<N/2$. All the data presented are free of finite-size effects.


The transform-limited optical pulse is defined by its mean energy
$\hbar \omega$ and the temporal width $\tau$ of the gaussian envelope
$\sim exp\{-(t/\tau)^2\}$.  We define an excitation energy $E_{exc}$
refered to the bottom of the (ordered) absorption band, i.e.
$E_{exc}=\hbar \omega - E_{gap}$.  All results are given for $\tau =
100$ fs, which corresponds to an energetic width (FWHM) of 22 meV.

To make contact with the previous work where two
particles in a single band were placed initially at a single site,
we first consider
the situation of a symmetric band structure with
$J^e = - J^h = 20$ meV. The absorption spectra with and without
Coulomb interaction are shown in Fig.~\ref{fig1} for the ordered case.
The peak structure near the absorption is due to the 
excitonic resonances.  Upon changing the sign of the Coulomb interaction, the
bound state
resonances are shifted from the bottom to the top of the absorption
spectrum.

As the dynamics of electrons and holes are the same for the assumed
symmetric band structure, we restrict our discussion to the electrons.
We first discuss the situation in the absence of Coulomb interaction.
Fig.~\ref{fig2} shows the corresponding $\Lambda_e(t)$ for 
two different
disorder parameters $W$ after excitation by a pulse at $E_{exc}=80$
meV. The excitation is centered in the absorption spectrum as
indicated in Fig.~\ref{fig1}. $\Lambda_e(t)$ evolves exponentially
with rise time less than 
1 ps.  Here and below, we take the
saturation value as a measure of localization.  As expected, it
decreases rapidly with increasing disorder.  We find $\Lambda \sim
W^{-1.3}$ as $W$ is varied over the range 40 meV to 240 meV for $J=20$
meV. A discussion of related exponents can be found in \cite{imre}.

Fig.~\ref{fig2} 
contrasts the interacting and noninteracting behavior for two
values of disorder and reveals three 
remarkable features. i) The interaction clearly leads to a
reduction of the localization of the particles.  We have carefully
checked that the saturation value of $\Lambda_e(t)$ at long times is
not due to a finite size effect; values $\leq N/2$ are fully converged
with respect to the sample size.  ii) While the 
participation number in the noninteracting situation 
evolves exponentially and
saturates quickly ($ < 1$
ps), the interacting wave packets 
evolve diffusively and
reach their saturation values at much
longer times.  
iii) The sign of the Coulomb interaction ($U=\pm 1$)
has virtually no influence on the propagation of the particles in this
case.  The same is true if we apply a very short excitation pulse
which spectrally covers the whole band. The spectral position of the
central pulse frequency within the band is then completely
irrelevant. In this situation the excited particle pair-wave packet is
initially situated exclusively at site $i=0$.  These observations are
not new \cite{Shepel,Imry,Jaquod}.  However, iii) has been
questioned \cite{frahm} and ii) remained unexplained.

In all cases where $J^e$ and $J^h$ are of comparable magnitude we find
that the participation number is enhanced by the interaction.  In a
mean field picture, it is the temporal fluctuations of the field
originating from the partner particle which destroy the coherence
necessary to produce localization.  This explanation in terms of a
{\it dynamic}-correlation-induced weakening of the influence of
disorder can be nicely corroborated by a number of case studies.

We note that contrary to previous statements, the independence
of the sign of the Coulomb interaction is not a general feature, but
is a consequence of the imposed electron-hole symmetry. In particular,
displacing the central frequency of excitation pulses from the center
of the absorption band, the situation changes completely. Note that
this choice of the excitation frequency corresponds to the realistic
situation where electron-hole pairs are excited close to the
absorption edge in semiconductors.  Fig.~\ref{fig3} shows the
participation number $\Lambda$ for light electrons and heavy holes,
i.e., $J^e = - 2 J^h = 20$ meV. The central excitation energy of the
pulse is placed in the lower part of the pair continuum at $E_{exc}
=40$ meV.  Results averaged over 60 realizations are shown for $W=80$
meV and $U=0, \pm 1$.  The results are invariant under reflection of
the excitation frequency through band center and simultaneous
switching of the sign of the interaction.  This 
reflects the approximate symmetry (within fluctuations in the site energy
distribution) of the Hamiltonian.

It is at first sight counterintuitive that the enhancement of the
participation number is larger for attractive ($U=-1$) than for
repulsive ($U=+1$) interaction. This behavior can be attributed to the
fact that for attractive interaction and positive masses (i.e. for
excitation into the lower half of the excitation continuum) the
electron-hole pair tends to stay closer together. The fluctuating
field due to the accompanying particle is then more pronounced as
compared to the case of repulsive interaction, where the mutually
repulsive particle pair tends to be separated. Hence the {\it
dynamic}-correlation-induced weakening of the influence of disorder is
less effective for repulsive than for attractive interaction.

Completely different behavior is found for a {\it static} field. We
consider an infinitely heavy hole, $J^h=0$, which now produces a
static field, and excitation at the (interaction-free) band
center. For both attractive and repulsive interactions, the
participation number is {\it decreased} with respect to the
noninteracting case. This result is easily understood without invoking
fluctuating fields since at band center electron states have maximal
extent.  In the presence of interaction, off-center states are admixed
leading to greater confinement. The effect of the static interaction
is thus opposite to that of a fluctuating field \cite{foot1}.

The strong retardation of the 
saturation in the interacting case can also be understood in our
picture.  Whether with or without interaction, the electron and hole
wave packets spread over a range given by the single-particle levels
involved in the optical transition just after the short excitation
pulse. The fluctuating Coulomb field due to the partner particle then
leads to an increase of the spread of the wave packets. As a
consequence, the average fluctuating field acting on a given particle
is reduced, which in turn tends to slow further spreading, eventually
leading to the observed saturation at long times.
The neglect of phonon interactions in our model is justified {\it 
a posteriori}. Fig.~\ref{fig2} makes it clear that 
the time scales between 100 fs and $\simeq$ 3 ps
are fully sufficient for experimental observations while near-band-edge acoustic phonon scattering occurs on longer time scales \cite{tak}.

Previous work \cite{Shepel,Imry,Jaquod} on the TIP problem suggests a
scaling of the two-particle localization length $l_2/l_1 \sim l_1^a
(U/J)^2$, with $a=1$ or $a=0.65$ \cite{frahm}.  Our results for the
participation number do not obey such a scaling law as far as the
dependence on $U$ is concerned.  We obtain for electrons and holes,
both for attractive and repulsive interaction, $\Lambda(U=\pm
1)/\Lambda(U=0) \sim \Lambda(U=0)^b$ with $b=0.65 \pm 0.3$.  
For electrons and attractive
interaction the present model predicts $\Lambda_e \sim W^{-c}$ with a
larger exponent $c \simeq 2.2$ compared to $c= 1.3$ for the
noninteracting case.

In conclusion, we have studied the localization of a pair of
interacting particles in a situation which, in principle, is
accessible to experiments.  Optical excitation in the pair continuum
of a disordered one-dimensional semiconductor with long-range Coulomb
interaction has been considered. Starting from a tight-binding
description, the temporal evolution of the participation numbers of
the electron and the hole wave packets has been calculated 
by a direct solution of the
equation of motion of the correlated material excitation
within linear response with regard to the exciting laser field. The
participation number increases with interaction for both attractive
and repulsive interaction. We find that in general the degree of
delocalization depends strongly on the sign of the interaction, in
contrast to previously published predictions.  The sign of the
interaction becomes irrelevant (even if the masses of electrons and
holes are different) only for two special situations: excitation in
the center of the pair continuum, or excitation of the whole band.  We
have checked that this result is independent of the assumed form of
the interaction and that it remains true also for the short-range
interactions studied in the literature. Compared to the single-band
models treated in the past, the present semiconductor model admits a
richer variety of phenomena, which can be qualitatively explained
within a mean-field picture. We emphasize that the enhancement of the
participation number is clearly not due to a finite size effect, and
that it should be experimentally observable.  Ultra-short
time-of-flight experiments on arrays of semiconductor quantum wires in
the coherent limit using pump-probe techniques are a promising
option. The enhancement should also be observable in disordered
semiconductor quantum wells. In this case we expect the enhancement to
be even more pronounced, since, in contrast with one-dimensional
systems, only states close to the band edge are essentially affected
by the disorder in two dimensions, so that the interaction 
will lead to coupling with rather extended states.

This work is supported by DFG, SFB 383 and 341, the Leibniz Prize,
OTKA (T021228, T024136, F024135), 
SNSF (2000-52183.97), and the A. v. Humboldt Foundation.
Discussions with A. Knorr and F. Gebhard are gratefully acknowledged.


\begin{figure}[ht]  
\begin{center}
\caption{Absorption spectrum 
of the ordered chain for $U=0, +1, -1$ and
for equal electron and hole masses 
($J^e = -J^h= 20$ meV). 
The spectrum  of the optical pulse with
$E_{exc}=80$ meV and $\tau = 100$ fs is also given.}
\label{fig1}
\end{center}
\end{figure}

\begin{figure}[ht]  
\begin{center}
\caption{Temporal evolution of the participation number $\Lambda$
after excitation by a 100 fs pulse
for the interacting (solid
lines) and noninteracting (dotted lines) cases. The disorder parameters
are $W=20$ meV and $W=60$ meV 
}

\label{fig2}
\end{center}
\end{figure}

\begin{figure}[ht]  
\begin{center}
\caption{Temporal evolution of $\Lambda_e$ (upper traces) and
$\Lambda_h$ (lower traces) for excitation in the lower half of the continuum
and $U=0, \pm 1$. 
$J^e = -2J^h= 20$ meV,
$W=80$ meV.}
\label{fig3}
\end{center}
\end{figure}


\begin{references}

\bibitem[*]{*} Permanent address:
Department of Theoretical Physics, Institute of Physics,
Technical University of Budapest, H-1520 Budapest, Hungary

\bibitem{dho} O. N. Dorokhov, Zh. Eksp. Teor. Fiz. {\bf 98}, 646  (1990)
		[Sov. Phys. JETP  {\bf 71}, 360  (1990)]. 

\bibitem{Shepel} D.L. Shepelyansky, 
                 Phys.~Rev.~Lett.~{\bf 73}, 2607 (1994).

\bibitem{Imry} Y. Imry, Europhys. Lett. {\bf 30}, 405 (1995).

\bibitem{Jaquod} Ph. Jacquod and D. L. Shepelyansky, 
                 Phys.~Rev.~Lett.~{\bf 75}, 3501 (1995);
                 Ph. Jacquod, D. L. Shepelyansky, and O.P. Sushkov, 
	         Phys.~Rev.~Lett.~{\bf 78}, 923 (1997).

\bibitem{Jaquod1}  Ph. Jacquod and D. L. Shepelyansky, 
Phys.~Rev.~Lett.~{\bf 78}, 4986 (1997).              

\bibitem{frahm} K. Frahm, A. M\"uller-Groeling, J.-L. Pichard,
and D. Weinmann, Europhys. Lett. {\bf 31}, 169 (1995).

\bibitem{evan} S. N. Evangelou and D. E. Katsanos,  
Phys.~Rev.~{\bf B56}, 12797 (1997).



\bibitem{frahm1} K. Frahm, A. M\"uller-Groeling, Europhys. Lett. {\bf 32},
385 (1995).

\bibitem{oppen} F. von Oppen, T. Wettig, J. M\"uller, Phys. Rev. Lett.
{\bf 76}, 491 (1996).

\bibitem{Roemer} R.A. R\"{o}mer, M. Schreiber, 
                 Phys.~Rev.~Lett.~{\bf 78}, 515 (1997).


\bibitem{Ponomarev} I.V. Ponomarev and P.G. Silvestrov, 
                    Phys.~Rev.~{\bf B56}, 3742 (1997).

\bibitem{foot} Tight-binding models can be successfully
applied to band-edge optics of
disordered semiconductors. See, e.g., S. Abe, Y. Toyozawa, Journ.
of Phys. Soc. Japan {\bf 50}, 2185 (1981);
Ch. Lonsky, P. Thomas, A. Weller, Phys. Rev. Lett. {\bf 63}, 652 (1989); 
One-dimensional models have been used, in particular, for numerical studies.
See, e.g., 
U. Dersch, M. Gr\"unewald, H. Overhof, P. Thomas, J. Phys. C:{\bf 20}, 121
(1987);
H. Overhof, K. Maschke, J. Phys. Cond. Matter {\bf 1}, 431 (1989);
K. Bott, S.W. Koch, P. Thomas,
Phys. Rev. {\bf B56}, 12784 (1997); D. Brinkmann, K. Bott, S.W. Koch,
P. Thomas, phys. stat. sol. (b) {\bf 206}, 493 (1998)


\bibitem{stein}  F. Steininger, A. Knorr, T. Stroucken, 
P. Thomas, S.W. Koch, Phys. Rev. Lett. {\bf 77}, 550 (1996).

\bibitem{stahl}  W. Huhn, A. Stahl, phys. stat. sol. (b) {\bf 124}, 167 (1984).




\bibitem{banyai} L. B\'anyai, I. Galbraith, C. Ell, H. Haug, 
Phys. Rev. {\bf B36}, 6099 (1987).




\bibitem{klaus} K. Bott, S.W. Koch, P. Thomas, Phys. Rev. {\bf B56}, 12784
(1997).

\bibitem{dirk} D. Brinkmann, K. Bott, S.W. Koch, P. Thomas,
phys. stat. sol. (b) {\bf 206}, 493 (1998)


\bibitem{haug} H. Haug and S.W. Koch, {\it Quantum Theory of the Optical and
Electronic Properties of Semiconductors}, (World Scientific, Singapore, 1990). 


\bibitem{wer}    K. Victor, V.M. Axt, A. Stahl, Phys. Rev.
{\bf B51}, 14164 (1995).

\bibitem{imre} I. Varga and J. Pipek, J. Phys.: Condens. Matter
{\bf 6}, L115 (1994).

\bibitem{foot1} We find that there is a gradual transition between both
regimes with a cross-over from static to dynamic behavior at
$J^h \sim J^e/5$.

\bibitem{tak} T. Takagahara, Phys. Rev. {\bf B31}, 6552 (1985);
D. Oberhauser, K.-H. Pantke, J.M. Hvam, G. Weimann, C. Klingshirn,
Phys. Rev. {\bf B47}, 6827 (1993);
J. Hoffmann, M. Umlauff, H. Kalt, phys. stat. sol. (b) {\bf 204}, 195 (1997).
\end{references}
\end{document}